\documentstyle[12pt,epsf]{article}

\title{CP Violation and Matter Effect in Long Baseline Neutrino
Oscillation Experiments}

\author{
 Jiro Arafune%
  \thanks{e-mail address: \tt arafune@icrr.u-tokyo.ac.jp},
 Masafumi Koike%
  \thanks{e-mail address: \tt koike@icrr.u-tokyo.ac.jp}
 and Joe Sato%
  \thanks{e-mail address: \tt joe@icrr.u-tokyo.ac.jp}\\
{\footnotesize \it
Institute for Cosmic Ray Research, University of Tokyo, Midori-cho,
Tanashi, Tokyo 188, Japan}
}

\date{}

\begin{document}
\maketitle

\begin{abstract}
    We show simple methods how to separate pure CP violating effect
    from matter effect in long baseline neutrino oscillation
    experiments with three generations of neutrinos.  We give compact
    formulae for neutrino oscillation probabilities assuming one of
    the three neutrino masses (presumably $\nu_{\tau}$ mass) to be
    much larger than the other masses and the effective mass due to
    matter effect.  Two methods are shown: One is to observe envelopes
    of the curves of oscillation probabilities as functions of
    neutrino energy; a merit of this method is that only a single
    detector is enough to determine the presence of CP violation.  The
    other is to compare experiments with at least two different
    baseline lengths; this has a merit that it needs only narrow
    energy range of oscillation data.
\end{abstract}

\section{Introduction}
The CP violation has been observed only in the hadron sector, and it
is very hard for us to understand where the CP violation originates
from.  If we observe CP violation in the lepton sector through the
neutrino oscillation experiments, we will be given an invaluable key
to study the origin of CP violation and to go beyond the Standard
Model.

The neutrino oscillation search is a powerful experiment which can
examine masses and/or mixing angles of the neutrinos.  The several
underground experiments, in fact, have shown lack of the solar
neutrinos \cite{Ga1,Ga2,Kam,Cl} and anomaly in the atmospheric
neutrinos \cite{AtmKam,IMB,SOUDAN2}\footnote{Some experiments have
not observed the atmospheric neutrino anomaly \cite{NUSEX,Frejus}.},
strongly indicating the neutrino oscillation \cite{Fogli1, FLM,
Yasuda}.  The solar neutrino deficit implies a mass difference of
$10^{-5} \sim 10^{-4} {\rm eV^2}$, while the atmospheric neutrino
anomaly suggests a mass difference around $10^{-3} \sim 10^{-2} {\rm
eV^2}$ \cite{Fogli1,FLM,Yasuda}.

The latter encourages us to make long baseline neutrino oscillation
experiments.  Recently such experiments are planned and will be
operated in the near future \cite{KEKKam,Ferm}. It is now desirable to
examine whether there is a chance to observe not only the neutrino
oscillation but also the CP or T violation by long baseline
experiments \cite{Tanimoto,ArafuneJoe}.  Two of the present authors
have studied how large T violation we may be seen in long baseline
experiments \cite{ArafuneJoe}, but they have not answered the question
how the CP violation is distinguished from the matter effect, in case
both the solar neutrino deficit and the atmospheric neutrino anomaly
are attributed to the neutrino oscillation (In this case the matter
effect is expected to give a fake CP-violation effect comparable to
pure CP-violation effect).

In this paper we will answer this question.  In sec. 2 we briefly
review neutrino oscillation for 3 generations, and give very compact
formulae describing neutrino oscillation in the presence of matter
(The detailed derivation of the formulae is given in the Appendix). In
sec. 3 we show two methods to distinguish pure CP violation from
matter effect.  In sec. 4 we summarize our work and give discussions.

\section{Compact Formulae for the Neutrino Oscillation Probabilities}
\subsection{Brief Review and Parameterization}
Let us briefly review CP violation in neutrino oscillation
\cite{FukugitaYanagida,BilenkyPetcov,Pakvasa} to clarify our notation.

We assume three generations of neutrinos which have mass eigenvalues
$m_{i} (i=1, 2, 3)$ and mixing matrix $U^{(0)}$ relating the flavor
eigenstates $\nu_{\alpha} (\alpha={\rm e}, \mu, \tau)$ and the mass
eigenstates in the vacuum $\nu\,'_{i} (i=1, 2, 3)$ as
\begin{equation}
  \nu_{\alpha} = U^{(0)}_{\alpha i} \nu\,'_{i}.
  \label{Udef}
\end{equation}
We parametrize $U^{(0)}$ \cite{ChauKeung,KuoPnataleone,Toshev} with
the Gell-Mann matrices $\lambda_{i}$'s as

\begin{eqnarray}
& &
U^{(0)}
=
{\rm e}^{{\rm i} \psi \lambda_{7}} \Gamma {\rm e}^{{\rm i} 
\phi \lambda_{5}} {\rm e}^{{\rm i} \omega \lambda_{2}} \nonumber 
\\
&=&
\left(
\begin{array}{ccc}
  1 & 0 & 0  \\
  0 & c_{\psi} & s_{\psi} \\
  0 & -s_{\psi} & c_{\psi}
\end{array}
\right)
\left(
\begin{array}{ccc}
  1 & 0 & 0  \\
  0 & 1 & 0  \\
  0 & 0 & {\rm e}^{{\rm i} \delta}
\end{array}
\right)
\left(
\begin{array}{ccc}
  c_{\phi} & 0 &  s_{\phi} \\
  0 & 1 & 0  \\
  -s_{\phi} & 0 & c_{\phi}
\end{array}
\right)
\left(
\begin{array}{ccc}
  c_{\omega} & s_{\omega} & 0 \\
  -s_{\omega} & c_{\omega} & 0  \\
  0 & 0 & 1
\end{array}
\right)
\nonumber \\
&=&
\left(
\begin{array}{ccc}
   c_{\phi} c_{\omega} &
   c_{\phi} s_{\omega} &
   s_{\phi}
  \\
   -c_{\psi} s_{\omega}
   -s_{\psi} s_{\phi} c_{\omega} {\rm e}^{{\rm i} \delta} &
   c_{\psi} c_{\omega}
   -s_{\psi} s_{\phi} s_{\omega} {\rm e}^{{\rm i} \delta} &
   s_{\psi} c_{\phi} {\rm e}^{{\rm i} \delta}
  \\
   s_{\psi} s_{\omega}
   -c_{\psi} s_{\phi} c_{\omega} {\rm e}^{{\rm i} \delta} &
   -s_{\psi} c_{\omega}
   -c_{\psi} s_{\phi} s_{\omega} {\rm e}^{{\rm i} \delta} &
   c_{\psi} c_{\phi} {\rm e}^{{\rm i} \delta}
\end{array}
\right),
\label{UPar2}
\end{eqnarray}
where $c_{\psi} = \cos \psi, s_{\phi} = \sin \phi$, etc.

The evolution equation for the flavor eigenstate vector in the vacuum
is
\begin{eqnarray}
 {\rm i} \frac{{\rm d} \nu}{{\rm d} x}
&=&
 - U^{(0)} {\rm diag}(p_1, p_2, p_3) U^{(0) \dagger} \nu \nonumber \\
&\simeq&
 \left\{
  - p_1 +
  \frac{1}{2 E}
  U^{(0)} {\rm diag} (0, \delta m^2_{21}, \delta m^2_{31})
  U^{(0) \dagger}
 \right\} \nu,
\end{eqnarray}
where $p_i$'s are the momenta, $E$ is the energy and $\delta m^2_{ij}
= m^2_i - m^2_j$.  Neglecting the term $p_1$ which gives an irrelevant
overall phase, we have
\begin{equation}
 {\rm i} \frac{{\rm d} \nu}{{\rm d} x}
=
 \frac{1}{2 E}
 U^{(0)} {\rm diag} (0, \delta m^2_{21}, \delta m^2_{31})
 U^{(0) \dagger}
 \nu.
 \label{VacEqnMotion}
\end{equation}

Similarly the evolution equation in matter is expressed as
\begin{equation}
 {\rm i} \frac{{\rm d} \nu}{{\rm d} x}
 = H \nu,
 \label{MatEqn}
\end{equation}
where
\begin{equation}
 H \equiv
 \frac{1}{2 E}
 U {\rm diag} (\mu^2_1, \mu^2_2, \mu^2_3) U^{\dagger},
 \label{Hdef}
\end{equation}
with a unitary mixing matrix $U$ and the effective mass squared
$\mu^{2}_{i}$'s $(i=1, 2, 3)$.  The matrix $U$ and the masses
$\mu_{i}$'s are determined by
\begin{equation}
U
\left(
\begin{array}{ccc}
  \mu^2_1 & 0 & 0  \\
  0 & \mu^2_2 & 0 \\
  0 & 0 & \mu^2_3
\end{array}
\right)
U^{\dagger}
=
U^{(0)}
\left(
\begin{array}{ccc}
  0 & 0 & 0  \\
  0 & \delta m^2_{21} & 0 \\
  0 & 0 & \delta m^2_{31}
\end{array}
\right)
U^{(0) \dagger}
+
\left(
\begin{array}{ccc}
  a & 0 & 0  \\
  0 & 0 & 0 \\
  0 & 0 & 0
\end{array}
\right).
\label{MassMatrixInMatter}
\end{equation}
Here
\begin{eqnarray}
 a &\equiv& 2 \sqrt{2} G_{\rm F} n_{\rm e} E \nonumber \\
   &=& 7.56 \times 10^{-5} {\rm eV^{2}}
       \frac{\rho}{\rm g\,cm^{-3}}
       \frac{E}{\rm GeV},
 \label{aDef}
\end{eqnarray}
where $n_{\rm e}$ is the electron density and $\rho$ is the matter
density.  The solution of eq.(\ref{MatEqn}) is then
\begin{equation}
 \nu (x) = S(x) \nu(0)
 \label{nu(x)}
\end{equation}
with
\begin{equation}
 S \equiv {\rm T\, e}^{ -{\rm i} \int_0^x {\rm d} s H (s) }
 \label{Sdef}
\end{equation}
(T being the symbol for time ordering), giving the oscillation
probability for $\nu_{\alpha} \rightarrow \nu_{\beta} (\alpha, \beta =
{\rm e}, \mu, \tau)$ at distance $L$ as
\begin{eqnarray}
 P(\nu_{\alpha} \rightarrow \nu_{\beta}; L)
&=&
 \left| S_{\beta \alpha} (L) \right|^2.
 \label{alpha2beta}
\end{eqnarray}
The oscillation probability for the antineutrinos $P(\bar\nu_{\alpha}
\rightarrow \bar\nu_{\beta})$ is obtained by replacing $a \rightarrow
-a$ and $U \rightarrow U^{\ast} ({\rm i.e.\,} \delta \rightarrow
-\delta)$ in eq.(\ref{alpha2beta}).
 
We assume in the following the matter density is independent of space
and time for simplicity, and have
\begin{equation}
 S(x) = {\rm e}^{ -{\rm i} H x}.
 \label{simpleS}
\end{equation}

\subsection{Approximation of the Oscillation Probability}
If we attribute both the solar neutrino deficit and the atmospheric
neutrino anomaly to the neutrino oscillation with MSW solution for the
solar neutrinos, we find most plausible solutions to satisfy $\delta
m^2_{21} \ll \delta m^2_{31}$ and $a \ll \delta m^2_{31}$
\cite{Fogli1,FLM}.  In the following we assume $a, \delta m^2_{21} \ll
\delta m^2_{31}$ \footnote{For the case $\delta m^2_{21} \ll a \ll
\delta m^2_{31}$ see ref.\cite{ArafuneJoe}.}.  This case is also
interesting when we consider the long baseline neutrino oscillation
experiments to be done in the near future \cite{KEKKam,Ferm}.

Decomposing $H = H_0 + H_1$ with
\begin{equation}
 H_0
=
 \frac{1}{2 E} U^{(0)}
 \left(
 \begin{array}{ccc}
  0 &   &  \\
    & 0 &  \\
    &   & \delta m^2_{31}
 \end{array}
 \right)
 U^{(0) \dagger}
 \label{H0def} \\
\end{equation}
and
\begin{equation}
 H_1
=
 \frac{1}{2 E}
 \left\{
  U^{(0)}
  \left(
  \begin{array}{ccc}
   0 &                 &  \\
     & \delta m^2_{21} &  \\
     &                 & 0
  \end{array}
  \right)
  U^{(0) \dagger} +
  \left(
  \begin{array}{ccc}
   a &   &  \\
     & 0 &  \\
     &   & 0
  \end{array}
  \right)
 \right\},
 \label{H1def}
\end{equation}
we treat $H_1$ as a perturbation and calculate eq.(\ref{simpleS})
up to the first order in $a$ and $\delta m^2_{21}$.  Defining
$\Omega (x)$ and $H_1 (x)$ as
\begin{equation}
 \Omega (x) = {\rm e}^{ {\rm i} H_0 x } S(x)
 \label{Omegadef}
\end{equation}
and
\begin{equation}
 H_1 (x) = {\rm e}^{ {\rm i} H_0 x} H_1 {\rm e}^{ -{\rm i} H_0 x},
 \label{H1(x)def}
\end{equation}
we have
\begin{equation}
 {\rm i} \frac{ {\rm d} \Omega}{ {\rm d} x }
=
 H_1 (x) \Omega (x)
 \label{Omegaeq}
\end{equation}
and
\begin{equation}
 \Omega (0) = 1,
 \label{OmegaInit}
\end{equation}
which give the solution\footnote{We note the eq.(\ref{OmegaApprox}) is
correct for a case where the matter density depends on $x$.}
\begin{eqnarray}
 \Omega (x)
&=&
 {\rm T \, e}^{ -{\rm i} \int_{0}^{x} {\rm d} s H_1 (s)  }
 \nonumber \\
&\simeq&
 1 - {\rm i} \int_{0}^{x} {\rm d} s H_1 (s).
 \label{OmegaApprox}
\end{eqnarray}
We note the approximation (\ref{OmegaApprox}) requires
\begin{equation}
 \frac{a x}{2 E} \ll 1 \quad {\rm and} \quad
 \frac{\delta m^2_{21} x}{2 E} \ll 1
 \label{AppCond}.
\end{equation}
The equations (\ref{Omegadef}) and (\ref{OmegaApprox}) give
\begin{equation}
 S(x) \simeq {\rm e}^{ -{\rm i} H_0 x } +
             {\rm e}^{ -{\rm i} H_0 x }
              ( -{\rm i} ) \int_{0}^{x} {\rm d} s H_1 (s).
 \label{S0+S1}
\end{equation}
We then obtain the oscillation probabilities $P(\nu_{\mu} \rightarrow
\nu_{\rm e})$, $P(\nu_{\mu} \rightarrow \nu_{\mu})$ and $P(\nu_{\mu}
\rightarrow \nu_{\tau})$ in the lowest order approximation as
\begin{eqnarray}
& &
 P(\nu_{\mu} \rightarrow \nu_{\rm e}; L)
=
 4 \sin^2 \frac{\delta m^2_{31} L}{4 E}
 c_{\phi}^2 s_{\phi}^2 s_{\psi}^2
 \left\{
  1 + \frac{a}{\delta m^2_{31}} \cdot 2 (1 - 2 s_{\phi}^2)
 \right\}
 \nonumber \\
&+&
 2 \frac{\delta m^2_{31} L}{2 E} \sin \frac{\delta m^2_{31} L}{2 E}
 c_{\phi}^2 s_{\phi} s_{\psi}
 \left\{
  - \frac{a}{\delta m^2_{31}} s_{\phi} s_{\psi} (1 - 2 s_{\phi}^2)
  +
 \frac{\delta m^2_{21}}{\delta m^2_{31}} s_{\omega}
    (-s_{\phi} s_{\psi} s_{\omega} + c_{\delta} c_{\psi} c_{\omega})
 \right\}
 \nonumber \\
&-&
 4 \frac{\delta m^2_{21} L}{2 E} \sin^2 \frac{\delta m^2_{31} L}{4 E}
 s_{\delta} c_{\phi}^2 s_{\phi} c_{\psi} s_{\psi} c_{\omega}
 s_{\omega},
 \label{mu2e}
\end{eqnarray}
\begin{eqnarray}
 P(\nu_{\mu} \rightarrow \nu_{\mu})
&=&
 1 +
     4 \sin^2 \frac{\delta m_{31}^2 \; L}{4E}
     c_{\phi}^2 s_{\psi}^2
     \left\{ ( c_{\phi}^2 s_{\psi}^2 - 1 ) +
      \frac{a}{\delta m_{31}^2} \cdot
      2 s_{\phi}^2 (1 - 2 c_{\phi}^2 s_{\psi}^2)
     \right\} \nonumber \\
 &+& 2 \frac{\delta m_{31}^2 \; L}{2E}
     \sin \frac{L \; \delta m_{31}^2}{2E}
     c_{\phi}^2 s_{\psi}^2
     \left\{
        \frac{a}{\delta m_{31}^2}
        s_{\phi}^2 (2 c_{\phi}^2 s_{\psi}^2 - 1)
     \right. \nonumber \\
 &+&
     \left.
      \frac{\delta m_{21}^2}{\delta m_{31}^2}
        (s_{\phi}^2 s_{\psi}^2 s_{\omega}^2
       + c_{\omega}^2 c_{\psi}^2
       - 2 c_{\delta} c_{\psi} c_{\omega} s_{\phi} s_{\psi}
           s_{\omega} )
     \right\}
 \label{mu2mu}
\end{eqnarray}
and
\begin{eqnarray}
 P(\nu_{\mu} \rightarrow \nu_{\tau})
 &=& 4 \sin^2 \frac{\delta m^2_{31} L}{4E}
     c_{\phi}^4 c_{\psi}^2 s_{\psi}^2
     \left(
      1 - \frac{a}{\delta m^2_{31}} \cdot 4 s_{\phi}^2
     \right)
 \nonumber \\
 &+&
  2 \frac{\delta m^2_{31} L}{2E}
  \sin \frac{\delta m^2_{31} L}{2E}
  c_{\phi}^2 c_{\psi} s_{\psi}
  \left[
     \frac{a}{\delta m_{31}^2}
     2 c_{\phi}^2 c_{\psi} s_{\phi}^2 s_{\psi}
  \right. \nonumber \\
 &-&
  \left.
     \frac{\delta m_{21}^2}{\delta m_{31}^2}
     \left\{
      (c_{\omega}^2 - s_{\omega}^2 s_{\phi}^2)
      c_{\psi} s_{\psi}
     +
      c_{\delta} (c_{\psi}^2 - s_{\psi}^2)
      s_{\phi} c_{\omega} s_{\omega}
    \right\}
  \right] \nonumber \\
 &+&
  4 \frac{\delta m^2_{21} L}{2E}
  \sin^2 \frac{\delta m^2_{31} L}{4E}
  s_{\delta} c_{\phi}^2 s_{\phi} c_{\psi} s_{\psi}
  c_{\omega} s_{\omega}.
 \label{mu2tau}
\end{eqnarray}
(Detailed derivation is presented in the Appendix).  Recalling that
$P(\bar\nu_{\alpha} \rightarrow \bar\nu_{\beta})$ is obtained from
$P(\nu_{\alpha} \rightarrow \nu_{\beta})$ by the replacements $a
\rightarrow -a$ and $\delta \rightarrow -\delta$, we have
\begin{eqnarray}
 \Delta P(\nu_{\mu} \rightarrow \nu_{\rm e})
&\equiv&
 P(\nu_{\mu} \rightarrow \nu_{\rm e}; L)
-
 P(\bar\nu_{\mu} \rightarrow \bar\nu_{\rm e}; L)
 \nonumber \\
&=&
 \Delta P_1(\nu_{\mu} \rightarrow \nu_{\rm e}) +
 \Delta P_2(\nu_{\mu} \rightarrow \nu_{\rm e}) +
 \Delta P_3(\nu_{\mu} \rightarrow \nu_{\rm e})
 \label{DeltaPsdef}
\end{eqnarray}
with
\begin{eqnarray}
 \Delta P_1(\nu_{\mu} \rightarrow \nu_{\rm e})
&=& 
 16 \frac{a}{\delta m^2_{31}} \sin^2 \frac{\delta m^2_{31} L}{4 E}
 c_{\phi}^2 s_{\phi}^2 s_{\psi}^2 (1 - 2 s_{\phi}^2),
 \label{DeltaP1def} \\
 \Delta P_2(\nu_{\mu} \rightarrow \nu_{\rm e})
&=& 
 -4 \frac{a L}{2 E} \sin \frac{\delta m^2_{31} L}{2 E}
 c_{\phi}^2 s_{\phi}^2 s_{\psi}^2 (1 - 2 s_{\phi}^2),
 \label{DeltaP2def}
\end{eqnarray}
and
\begin{eqnarray}
 \Delta P_3(\nu_{\mu} \rightarrow \nu_{\rm e})
&=&
 -8 \frac{\delta m^2_{21} L}{2 E}
 \sin^2 \frac{\delta m^2_{31} L}{4 E}
 s_{\delta} c_{\phi}^2 s_{\phi} c_{\psi} s_{\psi} c_{\omega}
 s_{\omega}.
 \label{DeltaP3def}
\end{eqnarray}
Similarly we obtain
\begin{eqnarray}
 & &
 \Delta P(\nu_{\mu} \rightarrow \nu_{\mu})
 \nonumber \\
 &=&
 16 \frac{a}{\delta m^2_{31}}
 \left[
  \sin^2 \frac{\delta m^2_{31} L}{4 E} -
  \frac{1}{4} \frac{\delta m^2_{31} L}{2 E}
   \sin \frac{\delta m^2_{31} L}{2 E}
 \right]
 c_{\phi}^2 s_{\phi}^2 s_{\psi}^2
 \left(
  1 - 2 c_{\phi}^2 s_{\psi}^2
 \right)
 \label{Deltamu2mu}
\end{eqnarray}
and
\begin{eqnarray}
 & &
 \Delta P(\nu_{\mu} \rightarrow \nu_{\tau})
 \nonumber \\
 &=&
 -32 \frac{a}{\delta m^2_{31}}
 \left[
  \sin^2 \frac{\delta m^2_{31} L}{4 E} -
  \frac{1}{4} \frac{\delta m^2_{31} L}{2 E}
   \sin \frac{\delta m^2_{31} L}{2 E}
 \right]
 c_{\phi}^4 s_{\phi}^2 c_{\psi}^2 s_{\psi}^2
 \nonumber \\
 &+&
 8 \frac{\delta m^2_{21} L}{2 E}
 \sin^2 \frac{\delta m^2_{31} L}{4 E}
 s_{\delta} c_{\phi}^2 s_{\phi} c_{\psi} s_{\psi} c_{\omega}
 s_{\omega}.
 \label{Deltamu2tau}
\end{eqnarray}
Here we make some comments.
\begin{enumerate}
\item $P (\nu_{\alpha} \rightarrow \nu_{\beta})$'s and $\Delta P
    (\nu_{\alpha} \rightarrow \nu_{\beta})$'s depend on $L$ and $E$ as
    functions of $L/E$ apart from the matter effect factor $a \, ( = 2
    \sqrt{2} G_{\rm F} n_{\rm e} E )$.
\item At least four experimental data are necessary to determine the
    function $\Delta P (\nu_{\mu} \rightarrow \nu_{\rm e})$, since it
    has four unknown factors: $\delta m^2_{31}, \delta m^2_{21},
    c_{\phi}^2 s_{\phi}^2 s_{\psi}^2 (1-2 s_{\phi}^2)$ and $s_{\delta}
    c_{\phi}^2 s_{\phi} c_{\psi} s_{\psi} c_{\omega} s_{\omega}$.  In
    order to determine all the mixing angles and the CP violating
    phase, we need to observe $P (\nu_{\mu} \rightarrow \nu_{\mu})$
    and $P (\bar\nu_{\mu} \rightarrow \bar\nu_{\mu})$ in addition.
\item $\Delta P (\nu_{\mu} \rightarrow \nu_{\mu})$ is independent of
    $\delta$ and consists only of matter effect term.
\end{enumerate}

\section{Separation of Pure CP Violating Effect from the Matter Effect}
Next we investigate how we can divide $\Delta P (\nu_{\mu} \rightarrow
\nu_{\rm e})$ into a pure CP-violation part and a matter effect part
\footnote{It is straightforward to extend the following arguments to
other processes like $\nu_{\mu} \rightarrow \nu_{\tau}$.  We present
the cases of $\nu_{\mu} \rightarrow \nu_{\rm e}$ and $\bar\nu_{\mu}
\rightarrow \bar\nu_{\rm e}$ as examples.}.  The terms $\Delta P_1
(\nu_{\mu} \rightarrow \nu_{\rm e})$ and $\Delta P_2 (\nu_{\mu}
\rightarrow \nu_{\rm e})$, which are proportional to ``$a$'', are due
to effect of the matter along the path.  The term $\Delta P_3
(\nu_{\mu} \rightarrow \nu_{\rm e})$, which is proportional to
$s_{\delta}$, is due to the pure CP violation (We simply call $\Delta
P_i (\nu_{\mu} \rightarrow \nu_{\rm e})$ as $\Delta P_i$ hereafter).
In the following we introduce two methods to separate the pure CP
violating effect $\Delta P_3$ from the matter effect $\Delta P_1 +
\Delta P_2$.

\subsection{Observation of Envelope Patterns}
\begin{figure}
 \unitlength=1cm
  \begin{picture}(15,5)
  \unitlength=1mm
  \centerline{
   \epsfysize=5cm
   \epsfbox{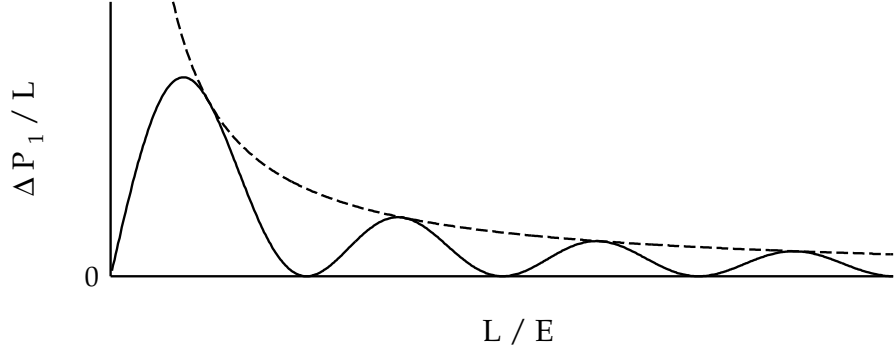}
   } 
\end{picture}
 \begin{flushleft}
     (a) Matter effect term $\Delta P_1(\nu_{\mu} \rightarrow \nu_{\rm
     e})$ divided by $L$ for $c_{\phi}^2 s_{\phi}^2 s_{\psi}^2 (1 - 2
     s_{\phi}^2) > 0$.  The envelope decreases monotonously with $L/E$.
 \end{flushleft}
 \label{x-1sinx}
 \unitlength=1cm
  \begin{picture}(15,5)
  \unitlength=1mm
  \centerline{
   \epsfysize=5cm
   \epsfbox{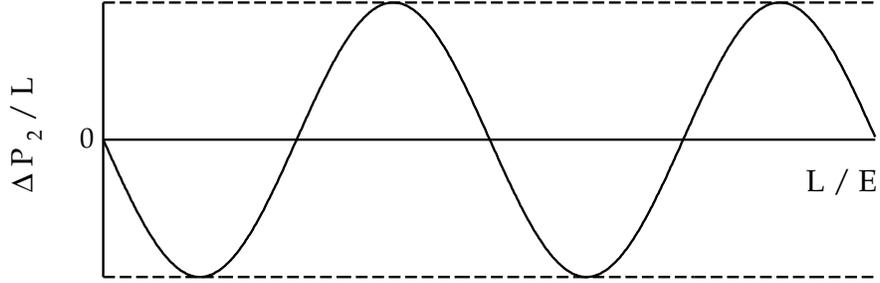}
   }
 \end{picture}
 \begin{flushleft}
     (b) Matter effect term $\Delta P_2(\nu_{\mu} \rightarrow \nu_{\rm
     e})$ divided by $L$ for $c_{\phi}^2 s_{\phi}^2 s_{\psi}^2 (1 - 2
     s_{\phi}^2) > 0$.  The envelope is flat.
 \end{flushleft}
 \label{sinx}
 \unitlength=1cm
  \begin{picture}(15,5)
  \unitlength=1mm
  \centerline{
   \epsfysize=5cm
   \epsfbox{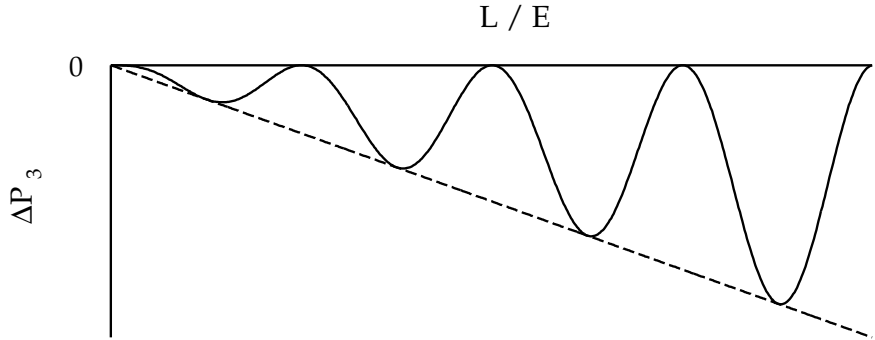}
   }
 \end{picture}
  \begin{flushleft}
   (c) CP-violation effect term
       $\Delta P_3(\nu_{\mu} \rightarrow \nu_{\rm e})$ for
       $ s_{\delta} c_{\phi}^2 s_{\phi} c_{\psi} s_{\psi} c_{\omega}
         s_{\omega} > 0$.  The envelope increases linearly with $L/E$.
  \end{flushleft}
 \label{xsinx}
 \caption{The oscillation behaviors of the $\Delta P_1, \Delta P_2$ and 
 $\Delta P_3$.}
 \label{OscBehave}
\end{figure}

One method is to observe the pattern of the envelope of $\Delta P$,
and to separate $\Delta P_3$ from it.  Considering the energy
dependence of $a (\propto E)$, we see that $\Delta P_1 /L$, $\Delta
P_2 /L$ and $\Delta P_3$ depend on a variable $L/E$ alone.  The
dependences of them on the variable $L/E$, however, are different from
each other as seen in Fig.  \ref{OscBehave}.  Each of them oscillates
with common zeros at $L/E = 2 \pi n / \delta m^2_{31} (n = 0, 1, 2,
\cdots)$ and has its characteristic envelope.  The envelope of $\Delta
P_1 /L$ decreases monotonously. That of $\Delta P_2 /L$ is flat.  That
of $\Delta P_3$ increases linearly.
It is thus possible to separate these three functions and determine CP
violating effect $\Delta P_3$ by measuring the probability $\Delta P$
over wide energy range in the long baseline neutrino oscillation
experiments.  This method has a merit that we can determine the pure
CP violating effect with a single detector.

\begin{figure}
 \unitlength=1cm
 \begin{picture}(13,9)
 \unitlength=1mm
 \put(106,78.5)  {$P(\nu_{\mu} \rightarrow \nu_{\rm e})$}
 \put(106,73)  {$P(\bar\nu_{\mu} \rightarrow \bar\nu_{\rm e})$}
 \put(106,67.5){$\Delta P$}
  \centerline{
   \epsfxsize=12.5cm
   \epsfbox{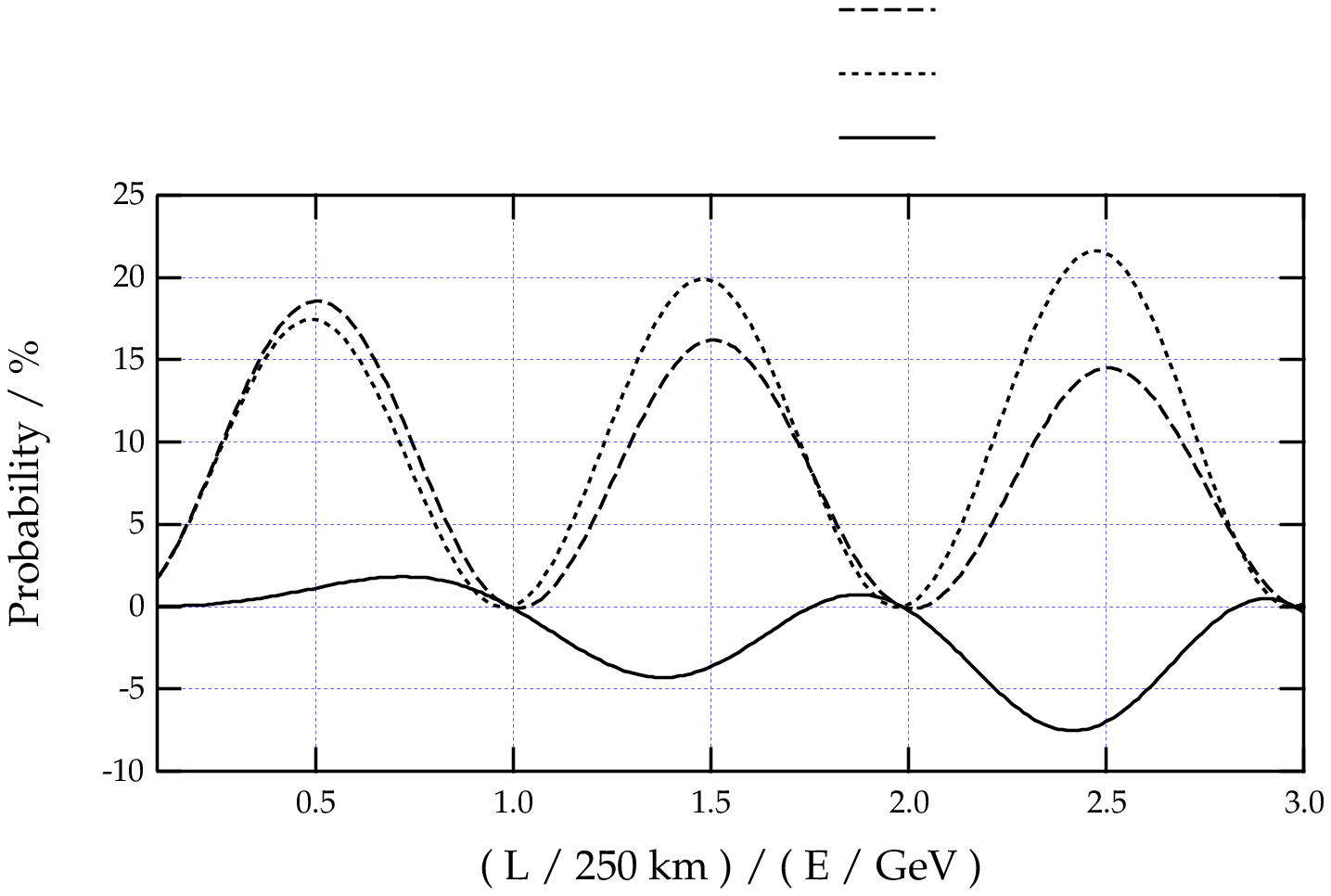}
   }
 \end{picture}
 \begin{center}
     (a) The oscillation probabilities as functions of $L/E$ for
     $\delta = \pi / 2$.
 \end{center}
 \label{mu2emax2}
\vspace{0.3cm}
 \begin{picture}(13,7)
 \unitlength=1mm
  \centerline{
   \epsfxsize=12.5cm
   \epsfbox{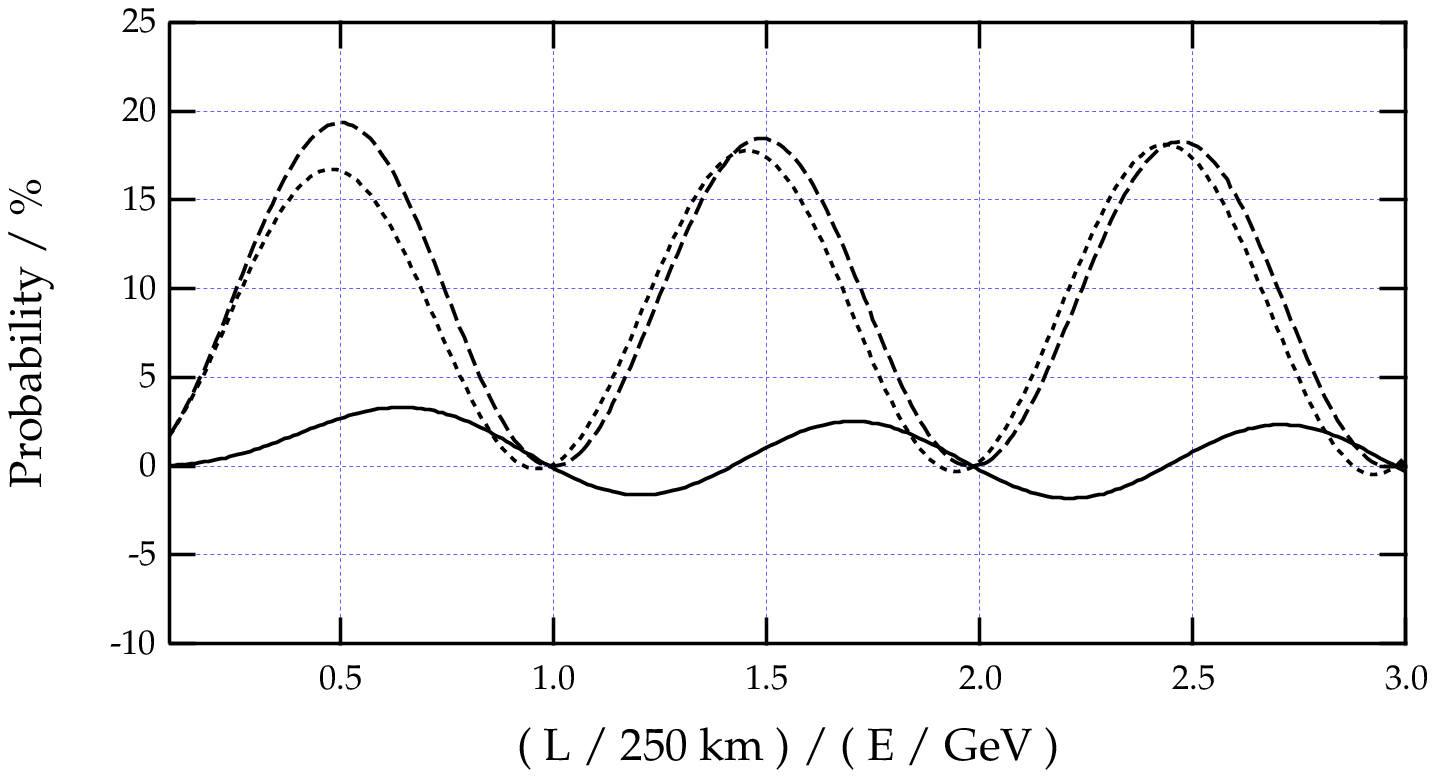}
   }
 \end{picture}
 \label{mu2e02}
 \begin{center}
     (b) The oscillation probabilities as functions of $L/E$ for
     $\delta = 0$.
 \end{center}
 \caption[OscProbCap]
 {The oscillation probabilities for $\delta = \pi / 2$
 (Fig. \ref{OscProb1}(a)) and $\delta = 0$ (Fig. \ref{OscProb1}(b)).
 $P(\nu_{\mu} \rightarrow \nu_{\rm e}), P(\bar\nu_{\mu} \rightarrow
 \bar\nu_{\rm e})$ and $\Delta P(\nu_{\mu} \rightarrow \nu_{\rm e})$
 are given by a broken line, a dotted line and a solid line,
 respectively.  Here $\rho = 3 {\rm g \, cm^{-3}}$ and $L=250 {\rm
 \,km}$ (the distance between KEK and Super-Kamiokande) are taken.  Other
 parameters are fixed at the following values which are consistent
 with the solar and atmospheric neutrino experiments \cite{FLM}:
 $\delta m^2_{21} = 10^{-4} {\rm \, eV^2}, \delta m^2_{31} = 10^{-2}
 {\rm \, eV^2}, s_{\psi} = 1/\sqrt{2}, s_{\phi} = \sqrt{0.1}$ and
 $s_{\omega} = 1/2$.}
 \label{OscProb1}
\end{figure}

\begin{figure}
 \unitlength=1cm
 \begin{picture}(15,10)
 \unitlength=1mm
 \put(122,68){$P(\nu_{\mu} \rightarrow \nu_{\rm e})$}
 \put(122,62){$P(\bar\nu_{\mu} \rightarrow \bar\nu_{\rm e})$}
 \put(122,56){$\Delta P$}
  \centerline{
   \epsfysize=8cm
   \epsfbox{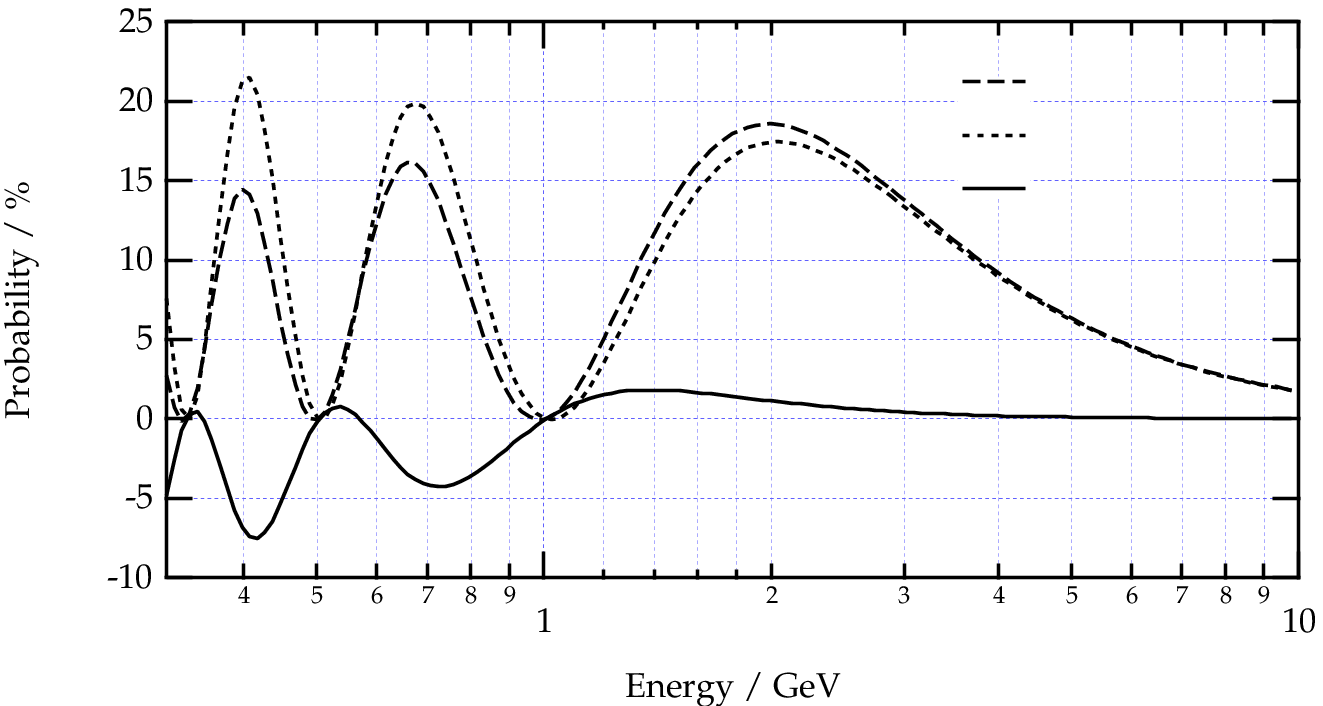}
   }
 \end{picture}
 \caption[OscProbCap]
 {The oscillation probabilities as functions of $E$.  Parameters are
 taken the same as in Fig. \ref{OscProb1}(a).}
 \label{OscProb2}
\end{figure}

In Fig. \ref{OscProb1} we give the probabilities $P(\nu_{\mu}
\rightarrow \nu_{\rm e})$ and $P(\bar\nu_{\mu} \rightarrow
\bar\nu_{\rm e})$ for a set of typical parameters which are consistent
with the solar and atmospheric neutrino experiments \cite{FLM}:
$\delta m^2_{21} = 10^{-4} {\rm \, eV^2}, \delta m^2_{31} = 10^{-2}
{\rm \, eV^2}, s_{\psi} = 1/\sqrt{2}, s_{\phi} = \sqrt{0.1}$ and
$s_{\omega} = 1/2$.  We see the effect of pure CP violation in
Fig. \ref{OscProb1}(a), since we find that the curve $\Delta P$ has the
envelope characteristic of $\Delta P_3$.
We show in Fig. \ref{OscProb2} the same probabilities as
Fig. \ref{OscProb1}(a) but as functions of $E$ to see the energy
dependence more directly.

We comment that the envelope behavior of $\Delta P$ can be understood
rather simply as follows: The term $\Delta P_3$ is proportional to
\cite{BWP,ArafuneJoe}
\begin{eqnarray}
 f
  &=&
 \sin \Delta_{21} + \sin \Delta_{32} + \sin \Delta_{13} 
 \nonumber \\
 &=&
 -4 \sin \frac{\Delta_{21}}{2} \sin \frac{\Delta_{32}}{2}
    \sin \frac{\Delta_{13}}{2},
 \label{fdef}
\end{eqnarray}
where
\begin{equation}
 \Delta_{ij} = \frac{ \delta \mu^2_{ij} L }{ 2E }.
 \label{Deltadef}
\end{equation}
Since we are interested in the first several peaks of $f$, we have
\begin{equation}
 \Delta_{31}, \Delta_{32} \sim {\cal O}(1).
 \label{order1}
\end{equation}
On the other hand we have
\begin{equation}
 \Delta_{21} \ll 1,
 \label{21small}
\end{equation}
because $\Delta_{21} \ll \Delta_{31}, \Delta_{32}$.  Taking into account
\begin{equation}
 \Delta_{21} + \Delta_{32} + \Delta_{13} = 0
 \label{DeltaSum}
\end{equation}
and eqs. (\ref{order1}) and (\ref{21small}), we obtain
\begin{equation}
 \Delta P_3 \propto f \simeq 2 \Delta_{21} \sin^2
 \frac{\Delta_{31}}{2}.
 \label{DeltaP3Behavior}
\end{equation}
This shows $\Delta P_3$ has a linearly increasing envelope
$\Delta_{21} \propto L/E$.  On the other hand, the envelopes of
$\Delta P_1$ and $\Delta P_2$ do not increase with $L/E$ for fixed
$L$, and it makes $\Delta P_3$ dominant in $\Delta P$ for large $L/E$.

\subsection{Comparison of Experiments with Different $L$'s}
The other method is to separate the pure CP violating effect by
comparison of experiments with two different $L$'s.  Suppose that two
experiments, one with $L = L_1$ and the other $L = L_2$, are
available.  We observe two probabilities $P (\nu_{\mu} \rightarrow
\nu_{\rm e}; L_1)$ at energy $E_1$ and $P (\nu_{\mu} \rightarrow
\nu_{\rm e}; L_2)$ at energy $E_2$ with $L_1 / E_1 = L_2 / E_2$.
Recalling that $P (\nu_{\mu} \rightarrow \nu_{\rm e}; L)$ is a
function of $L/E$ apart from the matter effect factor $a (\propto E)$,
we see that the difference
\begin{equation}
 \left\{
  P(\nu_{\mu} \rightarrow \nu_{\rm e}; L_1) -
  P(\nu_{\mu} \rightarrow \nu_{\rm e}; L_2)
 \right\}_{L_1/E_1 = L_2/E_2}
 \label{L1-L2}
\end{equation}
is due only to terms proportional to ``$a$''.  We obtain $\Delta P_3$
by subtracting these terms ($\Delta P_1 + \Delta P_2$) from $\Delta
P(\nu_{\mu} \rightarrow \nu_{\rm e})$ as\footnote{Note that the
eq.(\ref{SubMatterEff1}) does not require $P (\bar \nu_{\mu}
\rightarrow \bar \nu_{\rm e}; L_2)$.}
\begin{eqnarray}
& &
 \Delta P_{3} (\nu_{\mu} \rightarrow \nu_{\rm e}; L_1)
 \nonumber \\
&=&
 \left[
  \Delta P(\nu_{\mu} \rightarrow \nu_{\rm e}; L_1)
 -
  \frac{2 L_1}{L_2 - L_1}
  \left\{
   P(\nu_{\mu} \rightarrow \nu_{\rm e}; L_2) -
   P(\nu_{\mu} \rightarrow \nu_{\rm e}; L_1)
  \right\}
 \right]_{L/E={\rm const.}}
 \label{SubMatterEff1} \\
&=&
 \left[
  \Delta P(\nu_{\mu} \rightarrow \nu_{\rm e}; L_1)
 -
  \frac{L_1}{L_2 - L_1}
  \left\{
   \Delta P(\nu_{\mu} \rightarrow \nu_{\rm e}; L_2) -
   \Delta P(\nu_{\mu} \rightarrow \nu_{\rm e}; L_1)
  \right\}
 \right]_{L/E={\rm const.}}.
 \label{SubMatterEff2}
\end{eqnarray}
This method has a merit that it does not need to observe the envelope
nor many oscillation bumps in the low energy range.

\begin{figure}
 \unitlength=1cm
  \begin{picture}(15,11)
  \unitlength=1mm
 \put(52,103) {$P_{\rm KEK/SK} \equiv P(\nu_{\mu} \rightarrow \nu_{\rm e}; 
 L = 250 {\rm km})$}
 \put(52,97.5){$P_{\rm Minos} \equiv P(\nu_{\mu} \rightarrow \nu_{\rm e}; 
 L = 730 {\rm km})$}
 \put(52,92)  {$P_{\rm KEK/SK} - P_{\rm Minos}$}
 \put(52,86.5){CP violation $\Delta P_3$ for KEK/SK}
  \centerline{
   \epsfysize=11cm
   \epsfbox{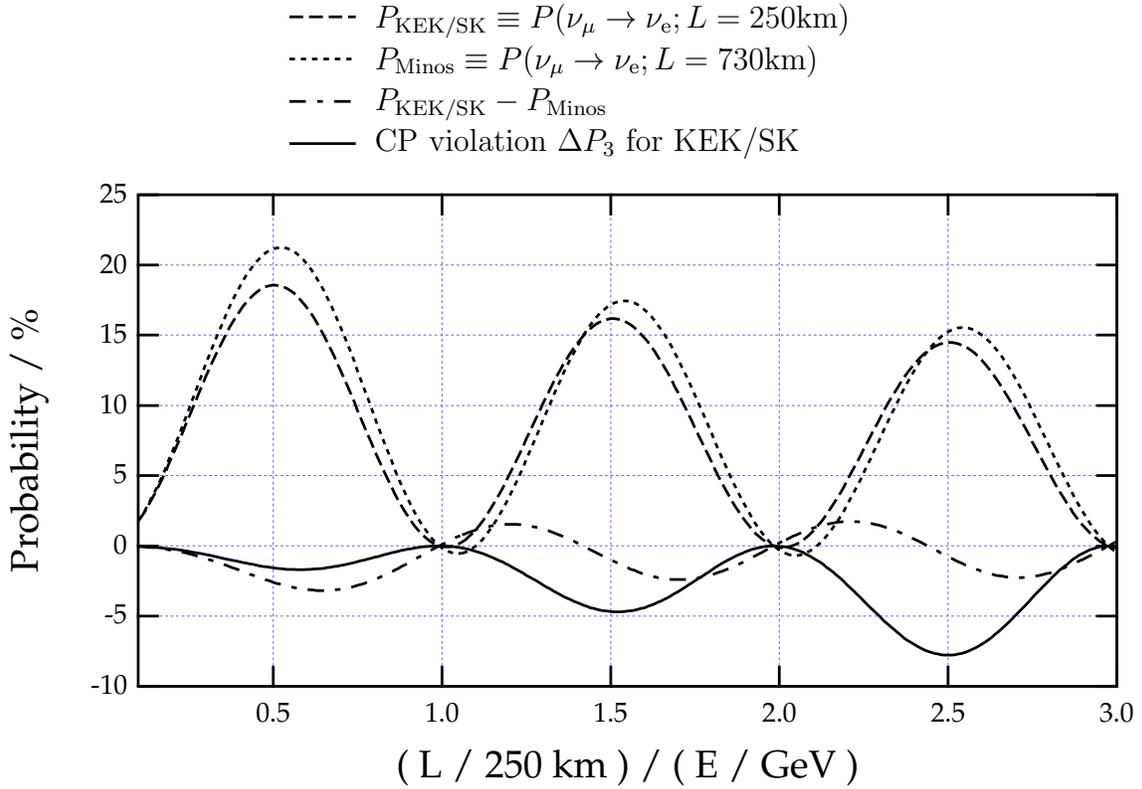}
   } 
\end{picture}
\caption[MinosKekCap]
{The oscillation probabilities $P(\nu_{\mu} \rightarrow \nu_{\rm
e})$'s for KEK/Super-Kamiokande experiment with $L=250 {\rm km}$
(broken line) and those for Minos experiment with $L=730 {\rm km}$
(dotted line).  Masses and mixing angles are the same as in Fig.
\ref{OscProb1}(a).  Their difference, which consists only of matter
effect, is shown by a dot-dashed line.  The pure CP violating effect
in KEK/Super-Kamiokande experiment determined by
eq.(\ref{SubMatterEff1}) is drawn by a solid line.}
 \label{MinosKek}
\end{figure}

In Fig. \ref{MinosKek} we compare $P(\nu_{\mu} \rightarrow \nu_{\rm
e})$ for $L=250 {\rm km}$ (KEK/Super-Kamiokande experiment) with that for
$L=730 {\rm km}$ (Minos experiment) in a case with the same neutrino
masses and mixing angles as those in Fig. \ref{OscProb1}(a) (or
Fig. \ref{OscProb2}) .  We see their difference, consisting only of the
matter effect, has the same shape as the solid line in Fig.
\ref{OscProb1}(b) up to a overall constant.  
We also show the pure CP violating effect obtained by the two
probabilities with eq.(\ref{SubMatterEff1}).  This curve
has a linearly increasing envelope as seen in Fig. \ref{OscBehave}(c).
\section{Summary and Discussions}

We have given very simple formulae for the transition probabilities of
neutrinos in long baseline experiments.  They have taken into account
not only the CP-violation effect but also the matter effect, and are
applicable to such interesting parameter regions that can explain both
the atmospheric neutrino anomaly and the solar neutrino deficit by the
neutrino oscillation.

We have shown with the aid of these formulae two methods to
distinguish pure CP violation from matter effect.  The dependence of
pure CP-violation effect on the energy $E$ and the distance $L$ is
different from that of matter effect: The former depends on $L/E$
alone and has a form $f(L/E)$, while the latter has a form $L \times
g(L/E) \equiv E \times \tilde{g}(L/E)$.  One method to distinguish is
to observe closely the energy dependence of the difference $
P(\nu_{\mu} \rightarrow \nu_{\rm e}; L) - P(\bar\nu_{\mu} \rightarrow
\bar\nu_{\rm e}; L) $ including the envelope of oscillation bumps.
The other is to compare results from two different distances $L_1$ and
$L_2$ with $L_1/E_1 = L_2/E_2$ and then to subtract the matter effect
by eq.(\ref{SubMatterEff1}) or eq.(\ref{SubMatterEff2}).

Each method has both its merits and demerits.  The first one has a
merit that we need experiments with only a single detector.  A merit
of the second is that we do not need wide range of energy (many bumps)
to survey the neutrino oscillation.

It is desirable to make long baseline neutrino oscillation experiments
with high intensity neutrino flux, and to study CP violation in the
lepton sector experimentally.

\appendix
\section{Derivation of the Oscillation Probabilities}
Here we present the derivation of eq.(\ref{mu2e}) $\sim$
eq.(\ref{mu2tau}) with use of eq.(\ref{S0+S1}), and show how well this
approximation works.
\begin{figure}
 \unitlength=1cm
  \begin{picture}(15,8)
  \unitlength=1mm
  \centerline{
   \epsfysize=8cm
   \epsfbox{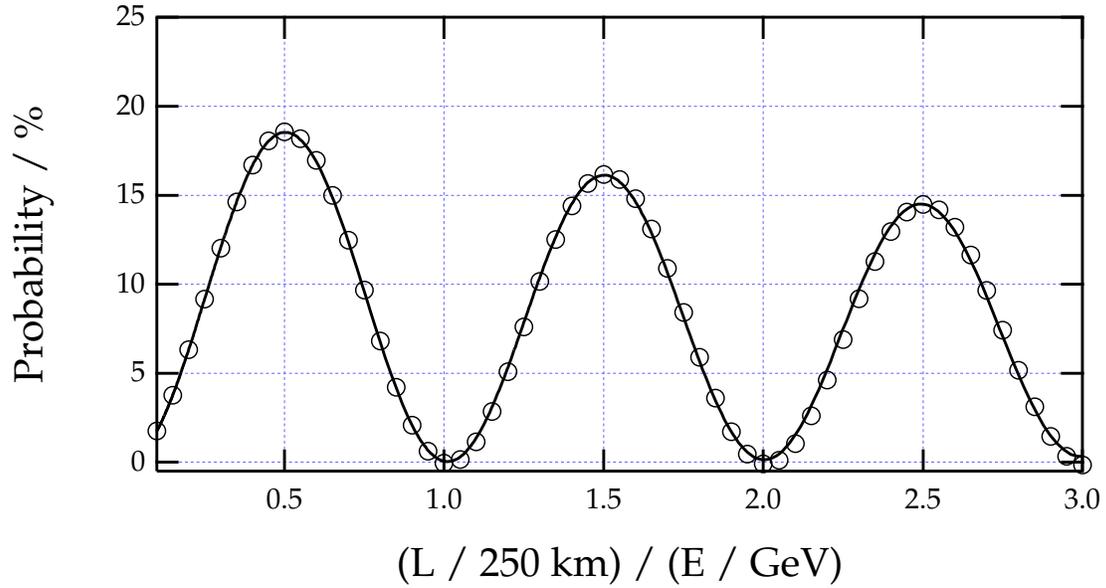}
   } 
\end{picture}
\begin{center}
    {(a) Exact and approximated values of $P (\nu_{\mu} \rightarrow
    \nu_{\rm e})$ for $L = 250$ km.}
\end{center}
\begin{picture}(15,8)
  \unitlength=1mm
  \centerline{
   \epsfysize=8cm
   \epsfbox{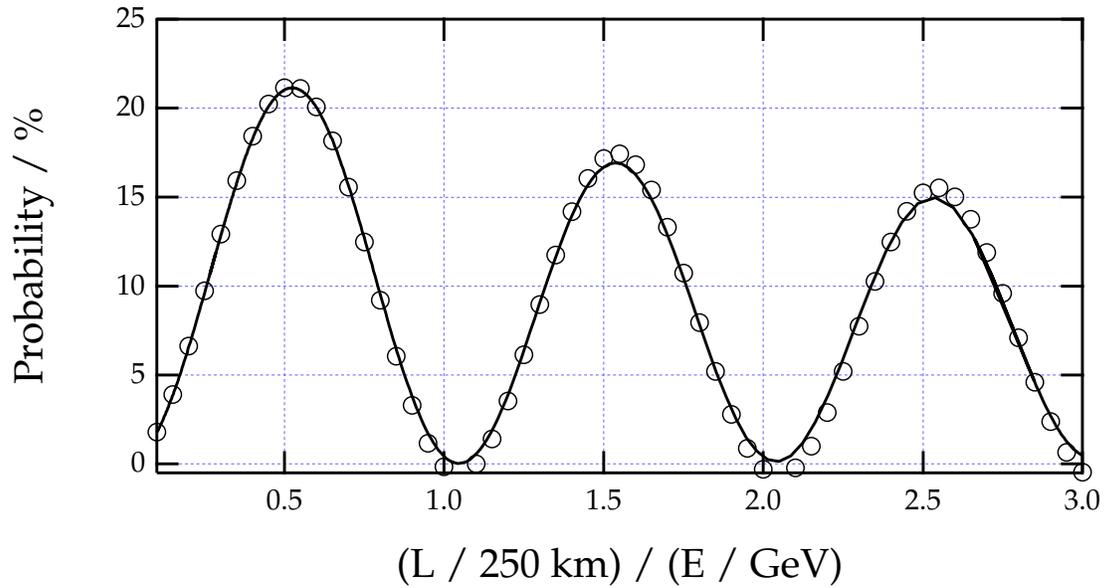}
   } 
\end{picture}
\begin{center}
    {(b) Exact and approximated values of $ P (\nu_{\mu} \rightarrow
    \nu_{\rm e})$ for $L = 730$ km.}
\end{center}
\caption[ExactApproxCompare]
{Exact and approximated values of $P (\nu_{\mu} \rightarrow \nu_{\rm
e})$ for $L = 250$ km (Fig. \ref{Exact&ApproxCompare}(a)) and those
for $L = 730$ km (Fig. \ref{Exact&ApproxCompare}(b)).  Exact values
and approximated ones are shown by solid lines and white circles,
respectively.  The parameters $s_{\psi}, s_{\phi}, s_{\omega}, \delta$
and $\rho$ are taken the same as in Fig. \ref{OscProb1}(a).}
 \label{Exact&ApproxCompare}
\end{figure}
Let us set $S(x) = S_0(x) + S_1(x)$, defining
\begin{eqnarray}
 S_0 (x) &=& {\rm e}^{-{\rm i} H_0 x},
 \label{S0def} \\
 S_1 (x) &=& {\rm e}^{-{\rm i} H_0 x}
             ({\rm -i}) \int_0^x {\rm d} s H_1 (s).
 \label{S1def}
\end{eqnarray}
We see 
\begin{eqnarray}
 S_0 (x)_{\beta \alpha}
&=&
 \left\{
  U^{(0)}
  {\rm e}^{-{\rm i} \frac{x}{2 E} {\rm diag}(0, 0, \delta m^2_{31})}
  U^{(0) \dagger}
 \right\}_{\beta \alpha}
 \nonumber \\
&=&
 \delta_{\beta \alpha} +
 U^{(0)}_{\beta 3} U^{(0) \ast}_{\alpha 3}
 ( {\rm e}^{ -{\rm i} \frac{\delta m^2_{31} x}{2 E}} - 1)
\end{eqnarray}
and
\begin{eqnarray}
 S_1 (x)_{\beta \alpha}
&=&
 - {\rm i} \int_0^x {\rm d}s \,
 \left[
  {\rm e}^{ -{\rm i} H_0 (x - s) } H_1 {\rm e}^{ -{\rm i} H_0 s }
 \right]_{\beta \alpha}
 \nonumber \\
&=&
  -{\rm i\,} U^{(0)}_{\beta i} U^{(0) \ast}_{\gamma i}
 (H_1)_{\gamma \delta} U^{(0)}_{\delta j} U^{(0) \ast}_{\alpha j}
 \Gamma(x)_{ij},
\end{eqnarray}
where
\begin{eqnarray}
 \Gamma(x)_{ij}
&\equiv&
 \int_0^x {\rm d}s \,
 {\rm e}^{ -{\rm i \,} \frac{\delta m^2_{31}}{2 E}
          \{ (x-s) \delta_{i3} + s \delta_{j3} \} }
 \nonumber \\
&=&
   \delta_{i3} \delta_{j3} \cdot
     x {\rm e}^{ -{\rm i} \frac{\delta m^2_{31} x}{2 E}}
  \nonumber \\
&+&
   \left\{
    (1 - \delta_{i3}) \delta_{j3} +
    \delta_{i3} (1 - \delta_{j3})
   \right\}
   \cdot
     \left( -{\rm i} \frac{\delta m^2_{31}}{2 E} \right)^{-1}
     \left( {\rm e}^{ -{\rm i} \frac{\delta m^2_{31} x}{2 E}} - 1
     \right)
  \nonumber \\
&+&
     (1 - \delta_{i3}) (1 - \delta_{j3}) \cdot x.
\end{eqnarray}
Using
\begin{eqnarray}
 U^{(0) \ast}_{\gamma i} (H_1)_{\gamma \delta} U^{(0)}_{\delta j}
&=&
 \frac{1}{2 E}
 \left\{
  {\rm diag}(0, \delta m^2_{21}, 0) +
  U^{(0) \dagger} {\rm diag}(a, 0, 0) U^{(0)}
 \right\}_{ij}
 \nonumber \\
&=&
 \frac{\delta m^2_{21}}{2 E} \delta_{i2} \delta_{j2} +
 \frac{a}{2 E} U^{(0) \ast}_{1i} U^{(0)}_{1j}
\end{eqnarray}
and
\begin{equation}
 \sum_{k=1}^2
 U^{(0) \ast}_{\alpha k} U^{(0)}_{1 k}
=
 \delta_{\alpha 1} -
 U^{(0) \ast}_{\alpha 3} U^{(0)}_{13},
\end{equation}
we obtain
\begin{equation}
 S(x)_{\beta \alpha}
 =
  \delta_{\beta \alpha} + {\rm i \,} T(x)_{\beta \alpha}
\end{equation}
with
\begin{eqnarray}
 {\rm i \,} T(x)_{\beta \alpha}
&=&
 -2 {\, \rm i\,} {\rm e}^{ -{\rm i} \frac{\delta m^2_{31} x}{4 E} }
 \sin \frac{\delta m^2_{31}}{4 E}
 U^{(0)}_{\beta 3} U^{(0) \ast}_{\alpha 3}
 \left[
   1
  -
   \frac{a}{\delta m^2_{31}}
  \left(
   2 \left| U^{(0)}_{13} \right|^2 - \delta_{\alpha 1} - \delta_{\beta
   1}
  \right)
  -
  {\rm i\,} \frac{a x}{2 E} \left| U^{(0)}_{13} \right|^2
 \right]
 \nonumber \\
&-&
 {\rm i\,} \frac{\delta m^2_{31} x}{2 E}
 \left[
   \frac{\delta m^2_{21}}{\delta m^2_{31}}
   U^{(0)}_{\beta 2} U^{(0) \ast}_{\alpha 2}
  +
 \right. \nonumber \\ & & \left.
   \frac{a}{\delta m^2_{31}}
   \left\{
    \delta_{\alpha 1} \delta_{\beta 1} \left| U^{(0)}_{13} \right|^2
   +
    U^{(0)}_{\beta 3} U^{(0) \ast}_{\alpha 3}
    \left(
    2 \left| U^{(0)}_{13} \right|^2 - \delta_{\alpha 3} -
    \delta_{\beta 3}
    \right )
   \right\}
 \right].
\end{eqnarray}
We then obtain the oscillation probability in the lowest order
approximation as
\begin{eqnarray}
& &
 P (\nu_{\alpha} \rightarrow \nu_{\beta}; L)
=
 \left| S(L)_{\beta \alpha} \right|^2
 \nonumber \\
&=&
 \delta_{\beta \alpha}
 \left[
  1
 -
  4 \left| U^{(0)}_{\alpha 3} \right|^2
  \sin^2 \frac{\delta m^2_{31} L}{4 E}
  \left\{
   1 - 2 \frac{a}{\delta m^2_{31}}
   \left(
   \left| U^{(0)}_{13} \right|^2 - \delta_{\alpha 1}
   \right)
  \right\}
 \right. \nonumber \\ & & \left.
 -
  2 \frac{a L}{2 E} \sin \frac{\delta m^2_{31} L}{2 E}
  \left| U^{(0)}_{\alpha 3} \right|^2
  \left| U^{(0)}_{13} \right|^2
 \right]
 \nonumber \\
&+&
 4 \left| U^{(0)}_{\beta  3} \right|^2
   \left| U^{(0)}_{\alpha 3} \right|^2
 \sin^2 \frac{\delta m^2_{31} L}{4 E}
 \left\{
  1
  -
  4 \frac{a}{\delta m^2_{31}} \left| U^{(0)}_{13} \right|^2
  +
  2 \frac{a}{\delta m^2_{31}} (\delta_{\alpha 1} + \delta_{\beta 1})
 \right\}
 \nonumber \\
&+&
 2 \frac{\delta m^2_{31} L}{2 E} \sin \frac{\delta m^2_{31} L}{2 E}
 \left[
  \frac{\delta m^2_{21}}{\delta m^2_{31}}
  {\, \rm Re\,}
  \left(
   U^{(0) \ast}_{\beta 3} U^{(0)}_{\beta 2}
   U^{(0)}_{\alpha 3} U^{(0) \ast}_{\alpha 2}
  \right)
 \right. \nonumber \\ & & \left.
 +
  \frac{a}{\delta m^2_{31}}
  \left\{
   \delta_{\alpha 1} \delta_{\beta 1} \left| U^{(0)}_{13} \right|^2
   +
   \left| U^{(0)}_{\alpha 3} \right|^2
   \left| U^{(0)}_{\beta 3} \right|^2
   \left(
    2 \left| U^{(0)}_{13} \right|^2
   -
    \delta_{\alpha 1} - \delta_{\beta 1}
   \right)
  \right\}
 \right]
 \nonumber \\
&-&
 4 \frac{\delta m^2_{21} L}{2 E}
 \sin^2 \frac{\delta m^2_{31} L}{4 E}
 {\,\rm Im \,}
 \left(
  U^{(0) \ast}_{\beta 3} U^{(0)}_{\beta 2}
  U^{(0)}_{\alpha 3} U^{(0) \ast}_{\alpha 2}
 \right).
 \label{alpha2betaapp}
\end{eqnarray}
Substituting eq.(\ref{UPar2}) in eq.(\ref{alpha2betaapp}) we finally
obtain eq.(\ref{mu2e}) $\sim$ eq.(\ref{mu2tau}).

Figure \ref{Exact&ApproxCompare} shows how well this approximation
works for KEK/Super-Kamiokande experiment and also for Minos
experiment with the same masses, mixing angles and CP violating phase
as in Fig. \ref{OscProb1}(a).  Our approximation requires (see
eq.(\ref{AppCond}))
\begin{equation}
 \frac{aL}{2E} =
 0.420 \left( \frac{L}{\rm 730 \,km} \right)
       \left( \frac{\rho}{\rm 3\,g\,cm^{-3}} \right)
 \ll 1
 \label{AppCond1}
\end{equation}
and
\begin{equation}
 \frac{\delta m^2_{21} L}{2E} =
 0.185
 \frac{(\delta m^2_{21} / {\rm 10^{-4} eV^2}) (L/{\rm 730 km})}
      {E / {\rm GeV}}
 \ll 1,
 \label{AppCond2}
\end{equation}
which is marginally satisfied for $L = 730 {\rm km}$.  We see that
even in this case eq.(\ref{alpha2betaapp}) gives good approximation.


\begin{thebibliography}{9}
\bibitem{Ga1} GALLEX Collaboration, P. Anselmann {\it et al.}, Phys.
    Lett. B {\bf 357}, 237 (1995).
    
\bibitem{Ga2} SAGE Collaboration, J.  N. Abdurashitov {\it et al.},
    Phys. Lett. B {\bf 328}, 234 (1994).
    
\bibitem{Kam} Kamiokande Collaboration, Y. Suzuki, Nucl. Phys. B (Proc.
    Suppl.) {\bf 38},54 (1995).
    
\bibitem{Cl} Homestake Collaboration, B. T. Cleveland {\it et al.},
    Nucl. Phys. B (Proc. Suppl.) {\bf 38}, 47 (1995).
    
\bibitem{AtmKam} Kamiokande Collaboration, K.S. Hirata {\it et al.},
    Phys.  Lett. B {\bf 205}, 416 (1988); {\it ibid.} B {\bf 280}, 146
    (1992) ; Y.  Fukuda {\it et al.}, Phys. Lett. B {\bf 335}, 237
    (1994).
    
\bibitem{IMB} IMB Collaboration,
    D. Casper {\it et al.}, Phys. Rev. Lett. {\bf 66}, 2561 (1991);\\
    R. Becker-Szendy {\it et al.}, Phys. Rev. D {\bf 46}, 3720 (1992).

\bibitem{SOUDAN2} SOUDAN2 Collaboration, T. Kafka, Nucl. Phys. B
    (Proc.  Suppl.) {\bf 35}, 427 (1994); M. C. Goodman, {\it ibid.}
    {\bf 38}, 337 (1995); W. W. M. Allison {\it et al.}, Phys. Lett. B
    {\bf 391}, 491 (1997).
    
\bibitem{NUSEX} NUSEX Collaboration, M. Aglietta {\it et al.},
    Europhys. Lett. {\bf 8}, 611(1989); {\it ibid.} {\bf 15}, 559
    (1991).
    
\bibitem{Frejus} Fr\'ejus Collaboration, K. Daum {\it et al.}, Z.
    Phys. C {\bf 66}, 417 (1995).
    
\bibitem{Fogli1} G.  L. Fogli, E. Lisi, D. Montanino and G.Scioscia,
    Preprint IASSNS-AST 96/41 (hep-ph/9607251).
    
\bibitem{FLM} G. L. Fogli, E.  Lisi and D. Montanino, Phys. Rev. D
    {\bf 49}, 3626 (1994).
    
\bibitem{Yasuda} O. Yasuda, preprint TMUP-HEL-9603 (hep-ph/9602342).

\bibitem{KEKKam} K. Nishikawa, INS-Rep-924 (1992).
    
\bibitem{Ferm} S. Parke, Fermilab-Conf-93/056-T (1993)
    (hep-ph/9304271).

\bibitem{Tanimoto} M. Tanimoto, Phys. Rev. D {\bf 55}, 322 (1997);
    M. Tanimoto, Prog. Theor. Phys. {\bf 97}, 901 (1997).
    
\bibitem{ArafuneJoe} J. Arafune and J. Sato, Phys. Rev. D {\bf 55},
    1653 (1997).
    
\bibitem{FukugitaYanagida} For a review, M. Fukugita and T. Yanagida,
    in {\it Physics and Astrophysics of Neutrinos}, edited by M.
    Fukugita and A. Suzuki (Springer-Verlag, Tokyo, 1994).

\bibitem{BilenkyPetcov} S. M. Bilenky and S. T. Petcov,
    Rev. Mod. Phys. {\bf 59}, 671 (1987).
    
\bibitem{Pakvasa} S. Pakvasa, in {\it High Energy Physics -- 1980},
    Proceedings of the 20th International Conference on High Energy
    Physics, Madison, Wisconsin, edited by L. Durand and L. Pondrom,
    AIP Conf. Proc. No. 68 (AIP, New York, 1981), Vol. 2, pp. 1164.

\bibitem{ChauKeung} L. -L. Chau and W. -Y. Keung, Phys. Rev. Lett.
    {\bf 59}, 671 (1987).

\bibitem{KuoPnataleone} T. K. Kuo and J. Pantaleone, Phys. Lett. B
    {\bf 198}, 406 (1987).

\bibitem{Toshev} S. Toshev, Phys. Lett. B {\bf 226}, 335 (1989).
    
\bibitem{BWP} V. Barger, K. Whisnant and R. J. N. Phillips, Phys. Rev.
    Lett. {\bf 45}, 2084 (1980).
\end{thebibliography}
\end{document}